\documentclass[twocolumn]{revtex4}
\usepackage{graphicx}
\begin{document}
\title{Photons in polychromatic rotating modes}
\author{S.J. van Enk$^{1,2}$ and G. Nienhuis$^3$}
\affiliation{$^1$Department of Physics, University of Oregon\\
Oregon Center for Optics and
Institute for Theoretical Science\\
 Eugene, OR 97403\\
 $^2$ Institute for Quantum Information\\
 California Institute of Technology\\
 Pasadena, CA 91125\\
$^3$Huygens Laboratorium, Universiteit Leiden\\
Postbus 9504, 2300 RA Leiden, The Netherlands}
\begin{abstract}
We propose a quantum theory of rotating light beams and study some
of its properties. Such beams are polychromatic and have either a slowly rotating
polarization or a slowly rotating transverse mode pattern. We show
there are, for both cases, three different natural types of modes
that qualify as rotating, one of which is a new type not
previously considered. We discuss differences between these three
types of rotating modes on the one hand and non-rotating modes as
viewed from a rotating frame of reference on the other. We present
various examples illustrating the possible use of rotating
photons, mostly for quantum information processing purposes.
We introduce in this context a rotating version of the two-photon singlet state.
\end{abstract}
\maketitle
\section{Introduction}
Several recent papers describe ``rotating beams of light''
\cite{soskin,alex,nien}. Such beams may have, for example, a
rotating linear or slightly elliptical polarization, and should
not be confused with circularly polarized light. At a fixed
instant of time the direction of the (linear or elliptical)
polarization vector rotates as a function of the propagation
coordinate, and in a fixed plane perpendicular to the propagation
direction the polarization rotates as a function of time. In a
different type of rotating light beams it is the transverse
intensity pattern rather than the polarization direction that is
rotating. Such rotating beams of light may be produced by passing
stationary beams through rotating optical elements, such as
astigmatic lenses and half-wave plates
\cite{garetz,kimble,niend,courtial}. The rotational frequency
$\Omega$ of the optical elements (and hence of the light beams) will
typically be very slow compared to the optical frequency $\omega$
of the light beam.

The theory discussed in the cited papers \cite{soskin,alex,nien}
is classical, although it is noted that some effects are more
conveniently understood in terms of photons. Some interesting
paradoxes and even some contradictions are mentioned in
\cite{soskin}, but the contradictions are not resolved there. 
The contradictions arise when one expects properties of rotating light beams
to equal those of nonrotating light beams as seen from a rotating frame of reference.
Here
we give a quantum description of rotating light beams. We show
that there are in fact several different natural ways of defining
``rotating photons''. Moreover, we show that a {\em different} type of rotating
photons arises by applying a rotation operator to standard
(non-rotating) quantized modes. The latter photons describe photons seen from a rotating frame. Carefully distinguishing these
different types of rotating photons thus removes the contradictions
mentioned above from Ref.~\cite{soskin}.

We will be particularly interested in the angular momentum of
rotating light beams. As we will show below, it will be easier to
calculate the average angular momentum in a quantum description
than in the classical descriptions of \cite{alex,soskin,nien}. For
example, it is pointed out in Refs.~\cite{alex,nien} that one
should be careful when applying expressions for the angular
momentum derived for {\em monochromatic} light beams
\cite{eg,berry}. Indeed, rotating light beams are necessarily {\em
polychromatic}. In the formalism we use here no such problems
arise and the quantum formalism takes care of polychromatic modes
automatically.

Naturally, most salient features of rotating photons can be
described in terms of angular momentum of light, simply because angular momentum operators generate rotations in space \cite{generator}.  This angular
momentum can have a spin or an orbital nature \cite{review}. We
thus start out by defining a complete set of electromagnetic field
modes as follows: we use monochromatic modes with definite values
of both spin and orbital angular momentum in the $z$ direction,
$S_z$ and $L_z$. The corresponding quantum numbers are denoted by
$\omega$ (for energy), $m$ (for orbital angular momentum) and
$s=\pm 1$ (for spin angular momentum, or, more precisely, for
helicity). The modes may be exact solutions of the Maxwell
equations (Bessel modes \cite{miceli}), or they may be exact
solutions of the paraxial equation, for modes propagating in the
$z$ direction (Laguerre-Gaussian modes). We must assume for the
exact Bessel modes, however, that they, too, are propagating
mostly in the $z$ direction. That is, we assume $k_T\ll \omega/c$,
with $k_T$ the magnitude of the {\em transverse} components of the
wave vector. This condition is needed in order for $L_z$ and $S_z$
to be well-defined angular momenta with integer eigenvalues
\cite{jmod}. So we use paraxial modes in either case.

A rotating mode or photon is then defined as an (almost) equal superposition of two opposite angular momenta $l$ and $-l$,
with different frequencies $\omega\pm l\Omega$. Photons with a rotating polarization are superpositions of two opposite spin angular momenta; photons with a rotating transverse mode pattern are superpositions of opposite orbital angular momenta. 

Besides the three quantum numbers mentioned so far, there is a
fourth quantum number necessary to fully specify an arbitrary
mode. This fourth quantum number describes the remaining
transverse spatial degree of freedom. It could be the number of
zeros $n_T$ in the transverse mode pattern of a Laguerre-Gaussian
mode or the transverse momentum $\hbar k_T$ of a Bessel mode
\cite{jmod}. For our purposes we do not have to specify the
transverse degrees of freedom any further. We thus assume that the
fourth quantum number is fixed, so that we can use a simplified
notation and denote the modes by the indices $(\omega,m,s)$.

The rest of the paper is organized as follows. In the next Section we will introduce notation and define more precisely modes with definite amounts of spin and orbital angular momenta. Those modes are monochromatic.  In Section \ref{Modes} we discuss how to define, in general, polychromatic modes. These are used to define quantized modes describing
rotating photons in Sections \ref{nonrot} and \ref{rot}. We will use the Heisenberg picture, as it allows
the most direct comparison of the fields with the classical case
treated before in the literature. As it turns out we can define at
least three different types of rotating modes, and we will discuss
the angular momentum of these various types of rotating photons.
In Section \ref{nonrot} we define rotating photons that have no angular momentum on average, in Section \ref{rot} we define two types of rotating photons with angular momentum, either parallel or anti-parallel to the propagation direction.
Measurements at the single-photon level of rotating modes are discussed at the end of Section \ref{nonrot}.
In Section \ref{frame} we define modes that correspond to nonrotating modes as seen from a rotating frame of reference, and we indicate the differences from the rotating modes of Sections \ref{nonrot} and \ref{rot}. In Section \ref{Quinfo} we
consider some applications of single-photon or two-photon states
of rotating modes, in particular, the use of rotating photons as a
means of encoding quantum information. We summarize in Section
\ref{sum}.
\section{Preliminaries}
For a given mode, the negative-frequency component of the
(dimensionless) classical electric field can be written in
cylindrical coordinates as
\begin{eqnarray}\label{Ec}
\vec{F}_{\omega,m,s}(z,\rho,\phi,t)=\exp(im\phi)\exp(-i\omega
t)F(\rho,z) \vec{e}_s,
\end{eqnarray}
which is valid for the free field. The polarization vectors are
$\vec{e}_{\pm}=(\vec{e}_{x}\pm i\vec{e}_{y})/\sqrt{2}$. There are
other nonzero components of the electric field but they are small
in the paraxial approximation. We focus our attention on the main
component (\ref{Ec}).

In Eq.~(\ref{Ec}) we left the dependence of the field on $\rho$
and $z$ unspecified. The precise form of $F$ depends on whether we
use exact or paraxial modes. For example, in the case of the exact
Bessel modes we have \cite{jmod}
\begin{equation}
F(\rho,z)\propto J_{|m|}(k_T\rho)\exp(ik_z z),
\end{equation}
where $k_z$ is the longitudinal component of the wave vector,
$k_z^2=k^2-k_T^2$, and $J_m$ is the $m$th-order Bessel function.
$F$ depends on the quantum numbers $k_T$ and $\omega$ in this
case, but not on $s$ and only on the absolute value  $|m|$. For
paraxial modes we have the more involved expressions for the
Laguerre-Gaussian modes \cite{allen}. Also in that case, $F$
depends on the quantum numbers $\omega$, the absolute value
$|m|$, and the quantum number $n_T$, but not on the polarization
index. This observation plays an important role later on, when we
define modes as superpositions of different modes that always have
the same value of $|m|$.

The Bessel modes do not diffract, and the transverse intensity
pattern is independent of $z$. There is a $z$-dependent phase
factor, and by choosing it equal to $\exp(ik_z (z-z_0))$ for some
fixed $z_0$ for {\em all} modes, we ensure that the mode functions
in the plane $z=z_0$ are independent of the frequency $\omega$, if
we fix the remaining quantum numbers $k_T,m,s$. We will refer to
the plane $z=z_0$ as the reference plane. For Bessel modes each
plane has the same intensity configuration. For paraxial modes, on
the other hand, we do have to define a particular location of the
reference plane, and we choose the same value $z_0$ for {\em all}
paraxial modes. Such modes form a complete set of (paraxial)
modes. Usually, the reference plane will be chosen as the focal
plane, where the wavefronts are flat.

\section{Time-dependent modes}\label{Modes}
\subsection{Field operators and mode functions}
Here we consider the theory of quantized modes, with as starting
point the mode functions (\ref{Ec}). We use the Heisenberg
picture, so that operators rather than states depend on time. The
(time-dependent) creation and annihilation operators for modes
with quantum numbers $\omega,m,s$ are denoted by
$\hat{a}^{\dagger}_{\omega,m,s}(t)$ and $\hat{a}_{\omega,m,s}(t)$.
(Recall that we leave out the transverse quantum numbers). The
frequency is a continuous variable, and the mode operators are
assumed to obey the standard bosonic commutation rules
$[\hat{a}_{\omega',m,s}(t),\hat{a}^{\dagger}_{\omega,m,s}(t)] =
\delta(\omega - \omega')$. Likewise, the single-photon states
$|\omega,m,s\rangle = \hat{a}^{\dagger}_{\omega,m,s} |{\rm
vac}\rangle$ are delta-function normalized. For a free field, the
time dependence of the mode operators is simply
\begin{equation}
\hat{a}_{\omega,m,s}(t)= \exp(-i\omega t)\hat{a}_{\omega,m,s}(0).
\end{equation}

We need the electric field operator from the relevant modes. In
the present paper, we can restrict ourselves to the paraxial modes
propagating in the positive $z$ direction. The contribution of a
single paraxial mode to the positive-frequency part of the
electric field operator is
    \begin{equation}\label{Ep}
\hat{\vec{E}}^{(+)}_{\omega,m,s}(t)=\sqrt{\frac{\hbar \omega}{4
\pi \epsilon_0 c}}\vec{F}_{\omega,m,s}\hat{a}_{\omega,m,s}(t)
=:\vec{E}_{\omega,m,s}\hat{a}_{\omega,m,s}(t)\;,
    \end{equation}
    with $\vec{F}$ given  by (\ref{Ec}).
Here we indicate operators by adorning them with hats.  The normalization
factor proportional to the square root of the frequency $\omega$
ensures the proper form of the Hamiltonian, in the form
    \begin{equation}
\hat{H} = \sum_{m,s} \int d\omega \;\hbar \omega
\;\hat{a}^\dagger_{\omega,m,s}\hat{a}_{\omega,m,s}\;.
    \end{equation}
    This
normalization is based on the assumption that the mode
functions $\vec{F}_{\omega,m,s}$ are normalized in each transverse
plane, as is common for paraxial modes. This means that $\int \rho
d\rho d\phi |\vec{F}_{\omega,m,s}|^2 = 1$.
The presence of the $\sqrt{\omega}$ term is responsible for the
existence of various different types of rotating photons, as we
will see below.

For later use we display the expression for contribution of a mode
to the positive-frequency part of the vector potential in the
Coulomb gauge as
    \begin{equation}\label{Ap}
\hat{\vec{A}}^{(+)}_{\omega,m,s}(t)=-i\sqrt{\frac{\hbar}{4 \pi
\epsilon_0 \omega c}}\vec{F}_{\omega,m,s}\hat{a}_{\omega,m,s}(t)
=:\vec{A}_{\omega,m,s}\hat{a}_{\omega,m,s}(t)\;.
    \end{equation}
Finally, we also give here the expression for the spin and orbital
angular momentum operators that will play a crucial role in the
rest of the paper:
\begin{eqnarray}
\hat{S}_z &=&\sum_{m,s} \int d\omega \;\hbar
s\;\hat{a}^\dagger_{\omega,m,s}
\hat{a}_{\omega,m,s}\;,\nonumber \\
\hat{L}_z &=&\sum_{m,s} \int d\omega \;\hbar
m\;\hat{a}^\dagger_{\omega,m,s} \hat{a}_{\omega,m,s}\;.
\end{eqnarray}
The operator for the total angular momentum is denoted as
$\hat{J}_z = \hat{S}_z + \hat{L}_z$.

\subsection{Unitary transformations of modes}
From now on we shall use for simplicity a generic subscript $i$ to
indicate the full set of quantum numbers $\omega$, $m$, $s$ and
the remaining transverse mode number, so that the summation over
$i$ represents a summation over $m$ and $s$ and an integration
over $\omega$. In this notation, the operators for the
positive-frequency part of the electric field and the vector
potential are denoted as
    \begin{equation}
\hat{\vec{E}}(t) = \sum_{i} \vec{E}_{i}\hat{a}_{i}(t)\;,\;
\hat{\vec{A}}(t) = \sum_{i} \vec{A}_{i}\hat{a}_{i}(t)\;.
\label{EAptot}
    \end{equation}
Given a complete set of field modes $\vec{E}_i$ and corresponding
mode operators $\hat{a}_i$, we can define a different complete set
of (orthonormal) modes and mode operators in a general way. For
this purpose we transform the field modes $\vec{E}_i$ as
\begin{equation}\label{UE}
\vec{E}'_i=\sum_j U_{ij}\vec{E}_j,
\end{equation}
where $U_{ij}$ is a unitary matrix. As will become obvious in the
next section, the specific transformations we will consider couple only a
limited number of modes, so that the summation in (\ref{UE})
extends over a few discrete indices $j$. A key feature of the
transformation is that it couples modes with different
frequencies, so that the primed modes are not monochromatic. In
order that the electric field operator can be expanded as
    \begin{equation}
\hat{\vec{E}}(t)=\sum_i\vec{E}'_{i}\hat{a}'_{i}(t)\;, \label{E'}
    \end{equation}
we must transform the set of annihilation operators by
\begin{equation}\label{Ua}
\hat{a}'_i(t)=\sum_j U^*_{ij}\hat{a}_j(t)\;.
\end{equation}
The unitarity of $U$ ensures that the new modes are still
orthogonal and that the new creation and annihilation operators
still satisfy the correct equal-time commutation relations. Due to
the unitarity of $U$, the inverse expansion of (\ref{Ua}) is given
by the transpose matrix, and we find for time zero
    \begin{equation}\label{Uainv}
\hat{a}_j(0)=\sum_i U_{ij}\hat{a}'_i(0)\;.
    \end{equation}

For the {\em free} field the time dependence of the electric-field
operator can be explicitly taken into account by incorporating the
time dependence in the mode functions. Substituting the inverse
expansion (\ref{Uainv}) into Eq.~(\ref{EAptot}) gives the
resulting expression
\begin{eqnarray}\label{trick}
\hat{\vec{E}}(t)&=&\sum_{i,j}
\vec{E}_j \exp(-i\omega_j t) U_{ij}\hat{a}'_i(0)\nonumber\\
&=:& \sum_i \vec{E}'_i(t)\hat{a}'_i(0)\;,
\end{eqnarray}
where the last line defines time-dependent and possibly
non-monochromatic mode functions $\vec{E}'_i(t)$. By this
transformation we can easily make a connection between the quantum
theory of rotating photons and the classical theory of rotating
light beams. On the basis of the new modes, the electric-field
operator $\hat{\vec{E}}(t)$ is now expressed either as an
expansion (\ref{E'}) with time-dependent mode operators, or as an
expansion (\ref{trick}) in time-dependent non-monochromatic mode
functions. It is noteworthy that although the summations are the
same, the summands are not. Only when the transformation does not
couple modes with different frequencies are the two expansions
(\ref{E'}) and (\ref{trick}) the same term by term.

\section{Rotation without angular momentum}\label{nonrot}
To describe modes rotating at a frequency $\Omega$ around the $z$
axis, we start with monochromatic modes at some frequency
$\omega$. We take equal superpositions of {\em two} fields with
opposite angular momenta (either spin or orbital) and shift their
frequencies by opposite amounts, proportional to the angular
momentum. In fact, we can choose to have a rotating transverse
intensity pattern by shifting the frequency proportional to the
orbital angular momentum, or a rotating polarization by shifting
in proportion to the spin angular momentum. It is convenient to
consider these cases separately.

\subsection{Rotating polarization}
As an example of Eq.~(\ref{Ua}), we define new mode operators
\begin{eqnarray}\label{rot1a}
\hat{b}_{\pm}:=
(\hat{a}_{\omega+\Omega s,m,s}\pm
\hat{a}_{\omega-\Omega s,m,-s})/\sqrt{2}.
\end{eqnarray}
This transformation is, by Eq.~(\ref{UE}), accompanied by the mode
definitions
\begin{equation}\label{rot1E}
\vec{E}^b_{\pm}:=
(\vec{E}_{\omega+\Omega s,m,s}
\pm
\vec{E}_{\omega-\Omega s,m,-s})/\sqrt{2}.
\end{equation}
For ease of notation we indicate the various transformed modes by
different letters, rather than by primes. The new modes $b$ are
described by new quantum numbers. For instance, $\omega$ is a
nominal frequency now, and no longer the eigenfrequency of the
mode. Indeed, the new mode has no eigenfrequency anymore. The
index $\pm$ replaces the bi-valued polarization index $s$. The
transverse quantum number, not displayed explicitly, stays the
same, and so does $m$, the orbital angular momentum. The rotation
frequency $\Omega$ is {\em not} an additional quantum number,  but
rather a parameter labeling the complete set of modes defined by
(\ref{rot1E}). Indeed, whereas fixing a particular quantum number
always restricts the set of modes to some smaller subset, fixing
$\Omega$ still leaves one with a complete set of modes. On the
other hand, we could try to consider $\Omega$ a quantum number if
we {\em fix} a particular value of $\omega=\omega_0$. This would
not be a natural choice, though, especially when
$\Omega\ll\omega$. Moreover, modes with frequencies larger than
$2\omega_0$ would not be included \footnote{On the flipside, our
mode definition requires that $\Omega<\omega$, so there is always
a range of frequencies $\omega$ for which rotating modes at
angular velocity $\Omega$ cannot be defined.}.

The reason for calling these new modes ``rotating'' is as follows:
the extra time-dependent terms in the electric field operator
rotating at a frequency $\pm \Omega$ due to the change in
frequency can be absorbed into the polarization part. For
instance, take $s=1$ and consider the reference plane
$z=z_0$. In that plane the mode functions $F(\rho,z_0)$ for the 2
modes appearing in the definition for $b$ are identical, by
construction. The transformation (\ref{trick}) for the
electric-field operators and the $b_{\pm}$ modes takes the form
\begin{equation}\label{trickb}
\vec{E}^b_{+}(t)\hat{b}_{+}(0) + \vec{E}^b_{-}(t)\hat{b}_{-}(0) =
\vec{E}^b_{+}\hat{b}_{+}(t) + \vec{E}^b_{-}\hat{b}_{-}(t)\;.
\end{equation}
In the reference plane these modes can be written as
\begin{eqnarray}\label{Eb}
\vec{E}^b_{\pm}(t)\hat{b}_{\pm} &=& \sqrt{\frac{\hbar\omega}{4 \pi
\epsilon_0 c}}\exp(-i\omega t)
\exp(im\phi)F(\rho,z_0)\hat{b}_{\pm}\nonumber\\
&\times&\left[
\cos\theta \vec{e}_+\exp(-i\Omega t)
\pm
\sin \theta
\vec{e}_-\exp(i\Omega t)
\right],\nonumber\\
\end{eqnarray}
where we define
\begin{eqnarray}\label{theta}
\cos\theta=\sqrt{\frac{\omega+\Delta}{2\omega}}\nonumber\\
\sin\theta=\sqrt{\frac{\omega-\Delta}{2\omega}},
\end{eqnarray}
with $\Delta=\Omega$ the frequency shift. The last line in
Eq.~(\ref{Eb}) is the time-dependent polarization vector
\begin{eqnarray}\label{pol}
\vec{e}(t) &=& A_{\pm}(\vec{e}_x\cos\Omega t +
\vec{e}_{y}\sin\Omega
t)\nonumber\\
&+& iA_{\mp}(-\vec{e}_x\sin\Omega t + \vec{e}_{y}\cos\Omega t)\;,
\end{eqnarray}
where
\begin{equation}
A_{\pm}=\frac{\cos\theta\pm\sin\theta}{\sqrt{2}}\;.
\end{equation}
Since typically we will have $\Omega\ll \omega$ (so that $A_+$ is
very close to $1$, and $A_-$ is close to $0$), the rotating
polarization is almost linear for both the $b_+$ mode and the
$b_-$ mode. Both modes $b_{\pm}$ describe an elliptical
polarization whose axes rotate in the same direction at a
frequency $\Omega$ around the $z$ axis. In fact, it is easy to
verify that a time shift by $\tau_1=\pi/(2\Omega)$ transforms the
$b_+$ mode into the $b_-$ mode and {\em vice versa}. More
precisely,
\begin{equation}
\vec{E}^b_{\pm}(t+\pi/(2\Omega))
=-i\exp(-i\pi\omega/(2\Omega))\vec{E}^b_{\mp}(t).
\end{equation}
Nevertheless these two modes are distinct, and the mode operators
$\hat{b}_+$ and $\hat{b}^{\dagger}_-$ commute.

So far we considered the field in the reference plane only. If we
go outside the reference plane $z=z_0$ then for Bessel modes we
still find the polarization is rotating in the same way in each
plane $z=$constant. For solutions of the paraxial equations
(Gaussian beams), however, the two components of the field at
different frequencies diffract in slightly different ways.
Nevertheless, as long as we do not stray too far from the
reference plane, the polarization still rotates in more or less
the same way. Similar conclusions will hold for all modes
discussed below. We will always display the field in the focal
plane $z=z_0$, and it should be kept in mind that our descriptions
are in general meant to apply near the reference plane.

\subsection{Rotating transverse mode pattern}
In a similar way we may define new modes by
\begin{eqnarray}\label{rot1m}
\hat{c}_{\pm}:=
(\hat{a}_{\omega+\Omega m,m,s}\pm
\hat{a}_{\omega-\Omega m,-m,s})/\sqrt{2}.
\end{eqnarray}
This transformation is, by Eqs.~(\ref{UE}) and (\ref{Ua}),
accompanied by the mode definitions
\begin{equation}\label{rot1Em}
\vec{E}^c_{\pm}:=
(\vec{E}_{\omega+\Omega m,m,s}
\pm
\vec{E}_{\omega-\Omega m,-m,s})/\sqrt{2}.
\end{equation}
These modes obey a relation similar as Eq.~(\ref{trickb}), with
$b$ replaced by $c$. Now the extra time-dependent terms in the
electric field operator rotating at a frequency $\pm \Omega m$ due
to the change in frequency can be absorbed into the azimuthal
part. Just as before, let us take $s=1$ and consider the reference
plane $z=z_0$. The electric-field operators for the $c_{\pm}$
modes in that plane can (using (\ref{trick})) be written as
\begin{eqnarray}\label{azi}
\vec{E}^c_{\pm}(t)\hat{c}_{\pm} &=& \sqrt{\frac{\hbar\omega}{4 \pi
\epsilon_0 c}}\exp(-i\omega t)
F(\rho,z_0)\vec{e}_+\hat{c}_{\pm}\nonumber\\
&\times&[
\cos\theta \exp(im(\phi-\Omega t))
\nonumber\\
&&\pm
\sin\theta
\exp(-im(\phi-\Omega t))
]
\end{eqnarray}
with the same definition (\ref{theta}) of the angle $\theta$ as
before, except that now $\Delta=m\Omega$. Clearly, the
time-dependent field has a transverse mode pattern that rotates
with a frequency $\Omega$ around the $z$ axis. Again, for both
modes $c_{\pm}$ the direction of the rotation is the same. And
just as for the polarization case, a time shift interchanges the
modes $c_{\pm}$. Now the time shift that accomplishes this is a
shift by $\tau_m=\pi/(2m\Omega)$, so that
\begin{equation}
\vec{E}^c_{\pm}(t+\pi/(2m\Omega))=-i
\exp(-i\pi\omega/(2m\Omega))\vec{E}^c_{\mp}(t).
\end{equation}
Finally, we note that the quantum numbers of the modes $c_{\pm}$
are different than those of the original modes $a$: in particular,
instead of the quantum number $m$ we have now both $m$ and $-m$,
while keeping $s$ and $\omega$ (although the latter is no longer
the eigenfrequency of the modes).

\subsection{Rotating polarization and mode pattern}
There is nothing to prevent us from defining modes where both the
transverse mode profile {\em and} the polarization are rotating at
a frequency $\Omega$. We just define
\begin{eqnarray}\label{rot1both}
\hat{d}_{\pm}:=
(\hat{a}_{\omega+\Omega (m+s),m,s}\pm
\hat{a}_{\omega-\Omega (m+s),-m,-s})/\sqrt{2}.
\end{eqnarray}
We can even define modes where the polarization is rotating at a
different frequency than the transverse mode pattern,
\begin{eqnarray}\label{rot1both2}
\hat{e}_{\pm}:=
(\hat{a}_{\omega+\Omega m+\Omega' s,m,s}\pm
\hat{a}_{\omega-\Omega m- \Omega' s,-m,-s})/\sqrt{2}.
\end{eqnarray}
Since all these redefinitions are unitary, the corresponding
electric field amplitudes will, by construction, still be valid
normalized solutions of the appropriate wave equations.

In order to see what it means to have the polarization  and the
transverse mode profile rotating at different frequencies, let us
consider one explicit example. Suppose we define
\begin{equation}
\hat{f}_{\pm}=(\hat{a}_{\omega+\Omega,+1,-1}\pm\hat{a}_{\omega-\Omega,-1,+1})/\sqrt{2}.
\end{equation}
Then this mode can be viewed as having a transverse mode pattern
that rotates in the positive direction at frequency $\Omega$.
Indeed, for any fixed linear polarization component, its field
distribution rotates in the positive direction. On the other hand,
the same mode can also be seen as having a polarization vector
that rotates in the {\em negative} direction. That is, if we fix
any point in the reference plane, then the local polarization vector rotates in
the negative direction. The reason is simple: the extra
time-dependent phase factors $\exp(\pm i\Omega t)$ can be absorbed
either in the polarization vector or in the transverse mode
pattern, and the choice is arbitrary, of course.
\begin{figure}
\begin{center}
\includegraphics[width=3.6in]{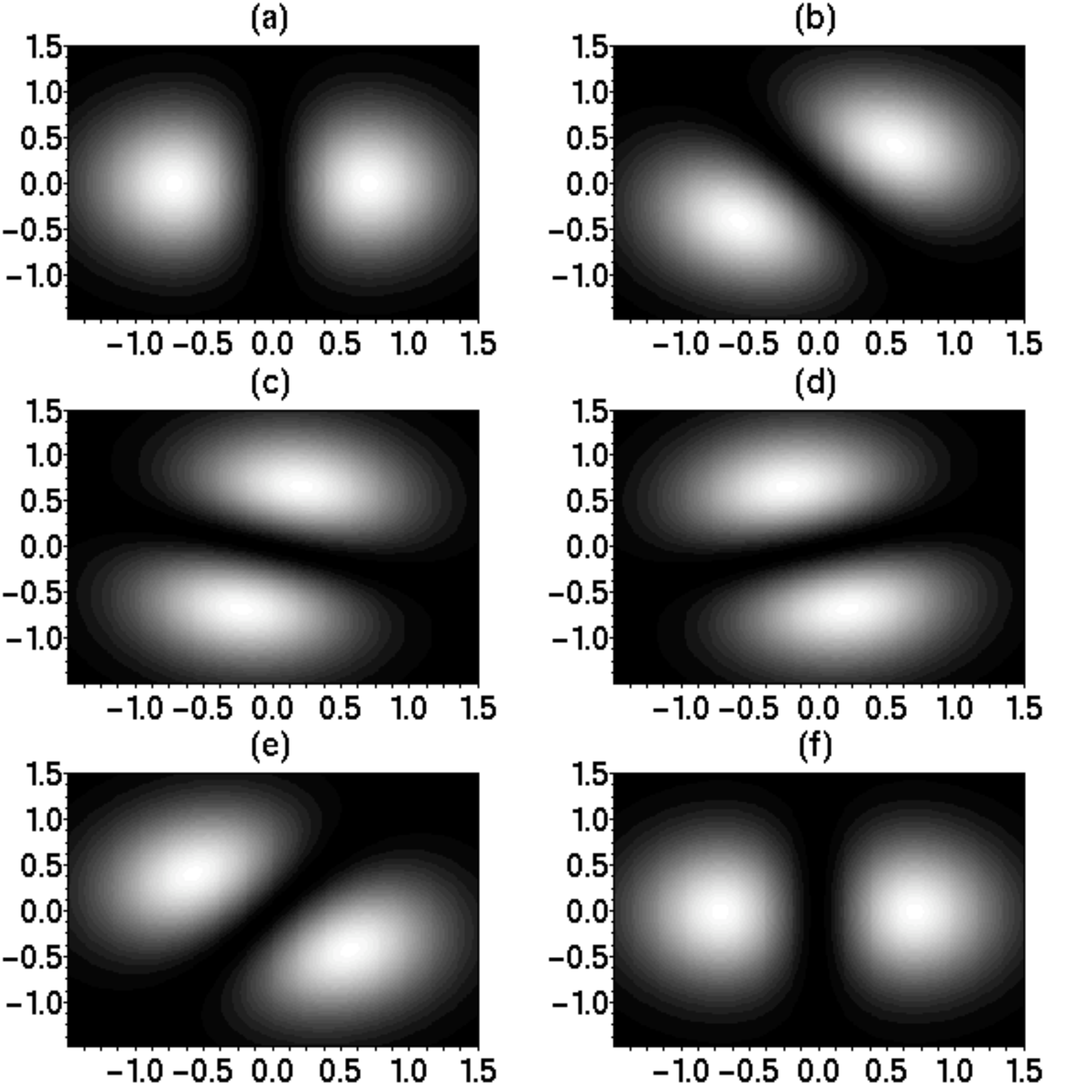}
\end{center}
\caption{Transverse intensity profile of the $x$ component of the field corresponding to the mode $f_+$, as a function of $x$ and $y$ (in arbitrary units) in the focal plane $z=0$. Snapshots are shown at 6 different times: for figures (a)--(f) we have $\Omega t=n\pi/5, n=0\ldots 5$, respectively. Here $\Omega>0$ and the sense of rotation of the intensity profile is positive.}
\end{figure}
\begin{figure}
\begin{center}
\includegraphics[width=3.6in]{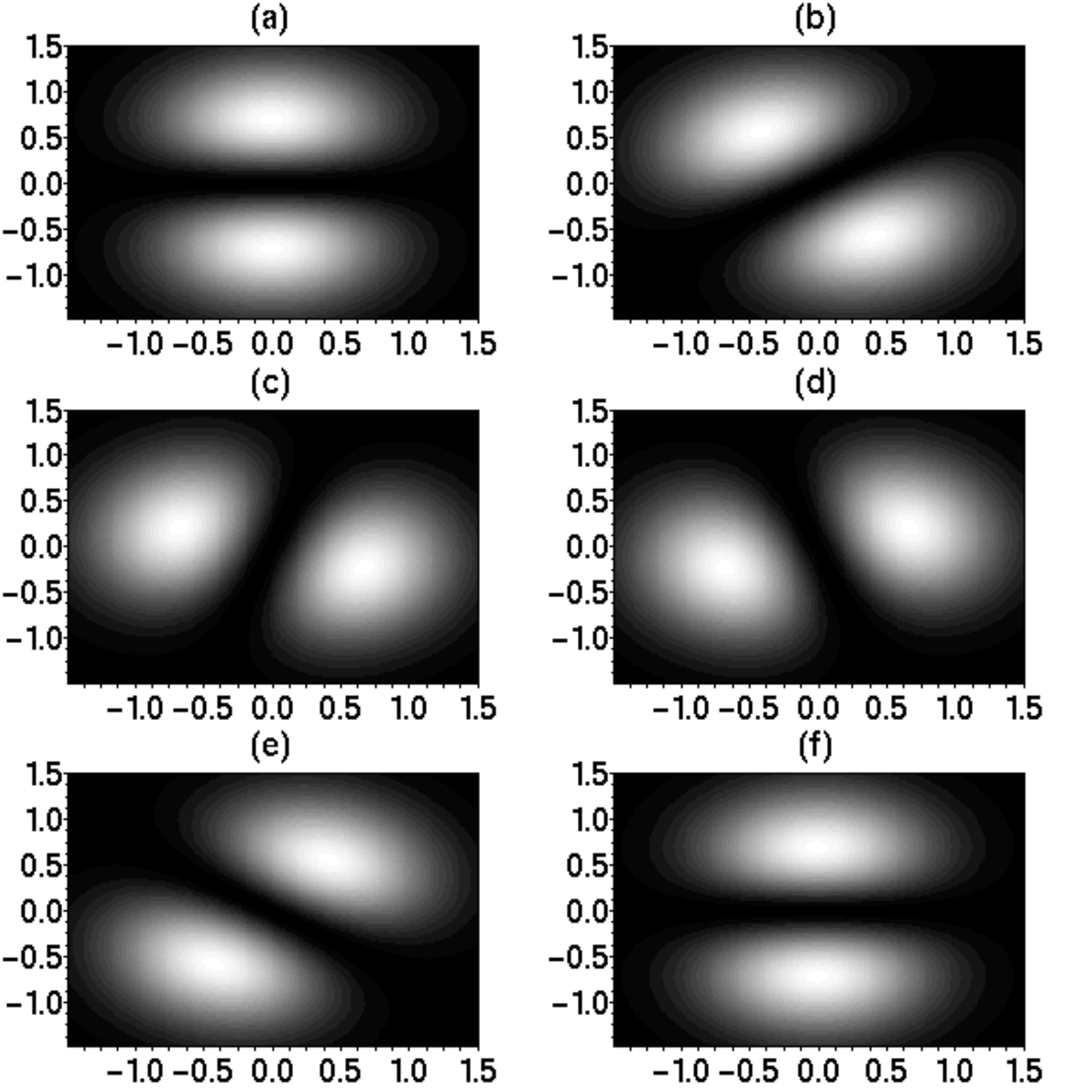}
\end{center}
\caption{Same as for Fig.~1, but for the $y$ component.}
\end{figure}
\begin{figure}
\begin{center}
\includegraphics[width=3.6in]{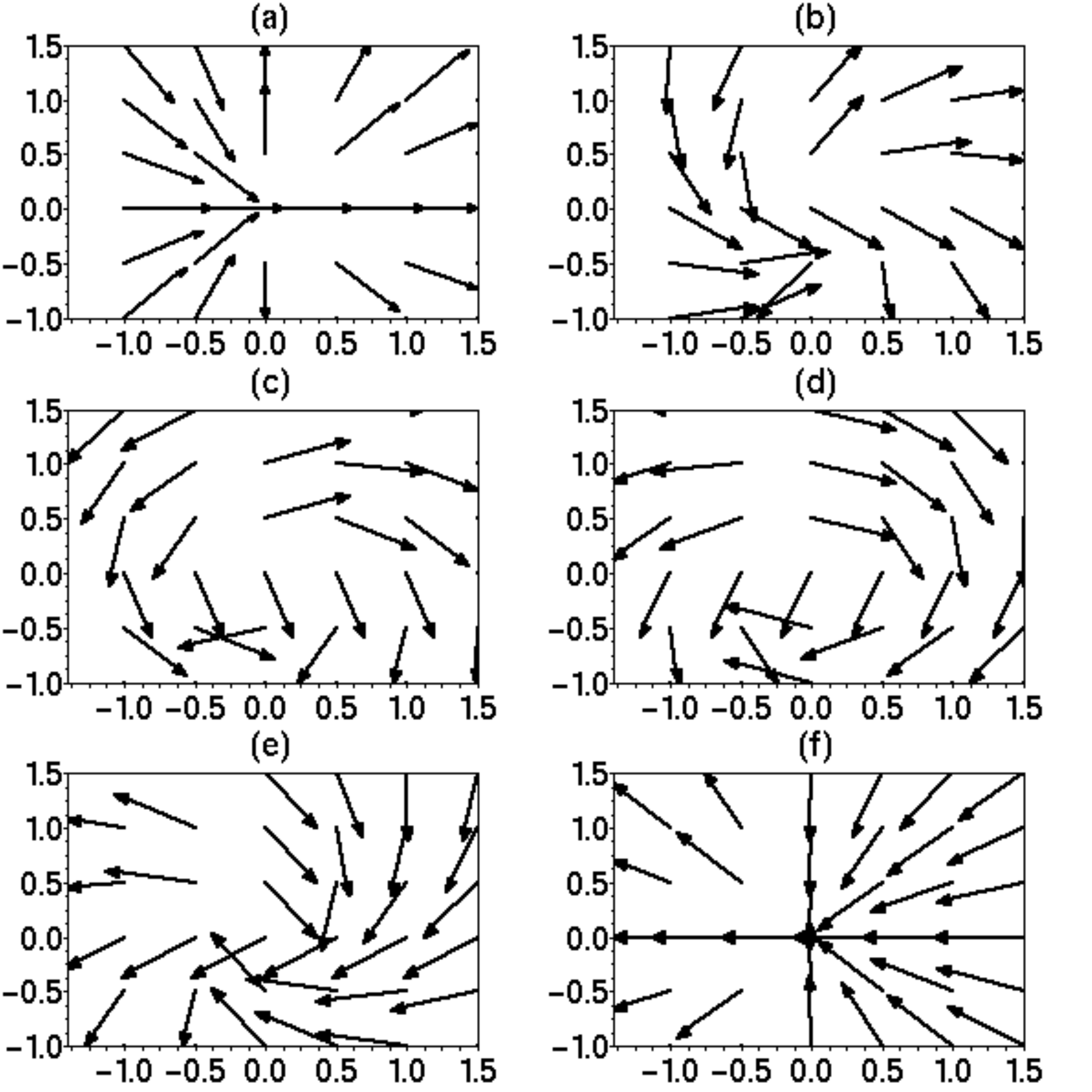}
\end{center}
\caption{The (linear) polarization vector of the field corresponding to the mode $f_+$. Snapshots are shown at the same instants as in Figs.~1 and 2.  Note the direction of rotation of the polarization vector is opposite (negative) to that of the transverse intensity patterns of Figs.~1 and 2, although we still have $\Omega>0$.}
\end{figure}
We illustrate this behavior in Figs.~1--3. We plot snapshots at different times of the intensity profiles for the $x$ and $y$ components of the field in Figs.~1 and 2, respectively. The polarization direction is plotted in Fig.~3 at the same instants of time.
\subsection{Rotating single photons produced by rotating mode inverters}
Now consider a single photon in any one of the modes we have
defined so far. For instance, consider a state of the form
\begin{equation} \label{1b}
|1\rangle_b=\hat{b}_+^\dagger|{\rm vac}\rangle,
\end{equation}
where $|{\rm vac}\rangle$ denotes the vacuum state, with all modes
unoccupied by photons. The coherence properties of a single-photon
state are characterized by the complex matrix element of the
electric-field operator
\begin{equation}
\langle {\rm vac}|\hat{\vec{E}}(\vec{r},t) |1\rangle_b =
\vec{E}_+^b(\vec{r},t)\;.
\end{equation}
This quantity, which is the quantum analog of the classical
electric field, is the detection amplitude function of the photon.
It determines the second-order coherence of a one-photon field
\cite{mandel}.

This photon has an average spin angular momentum of zero, although
its polarization is rotating at a frequency $\Omega$, according to
(\ref{pol}). The simple reason is that the photon is in an {\em
equal} superposition of spin angular momentum eigenstates with
eigenvalues  $+\hbar$ and $-\hbar$. There is no contradiction in
having a rotating polarization and yet zero spin, as there is no
simple {\em linear} relation between polarization and spin angular
momentum. The expectation value of the electric-field amplitude is
in fact always zero for any single-photon state, but the spin
angular momentum is determined, in both the classical case and the
quantum case, by a bilinear function of the field amplitudes; the
spin is nonzero for any single photon with a definite polarization
that is not linear. The average energy $\langle E\rangle$ of the
$b$ photon is $\hbar\omega$, again because it is in an {\em equal}
superposition of energy eigenstates with energies $\hbar\omega \pm
\hbar\Omega$.

Similar conclusions hold for the other modes, $c$, $d$, $e$, and
$f$ defined in the previous subsections. That is, for each such
mode the average energy of a single photon is $\hbar\omega$. For
the mode $c$, the orbital angular momentum vanishes, while for the
other modes the total angular momentum is zero, even though these
modes obviously display rotation.

In \cite{nien} it was shown that rotating photons are generated by
rotating mode inverters. Suppose one has an optical element that
``inverts'' the polarization vector of a light beam according to
\begin{equation}\label{map}
\vec{e}_{s}\mapsto\vec{e}_{-s}
\end{equation}
for $s=\pm 1$. This is the effect of a half-wave plate. Then a
plate that rotates at an angular frequency $\Omega/2$ will
generate a mode with a polarization vector rotating at a frequency
$\Omega$. The doubling of the rotation frequency can be understood
by noting that in a rotating frame the mapping (\ref{map}) becomes
\begin{equation}
\vec{e}_{s}\exp(is\Omega t/2)\mapsto\vec{e}_{-s}\exp(-is\Omega
t/2)\;.
\end{equation}
The quantum equivalent of this mapping is
\begin{equation}\label{Hmap}
\hat{a}_{\omega,m,s}\mapsto \hat{a}_{\omega-s\Omega,m,-s}\;.
\end{equation}
The linear superposition $(\hat{a}_{\omega,m,1} +
\hat{a}_{\omega,m,-1})/\sqrt{2}$ of mode operators corresponds to
a mode with linear polarization. This linear superposition is
mapped onto
\begin{equation}\label{bmap}
(\hat{a}_{\omega,m,1} + \hat{a}_{\omega,m,-1})/\sqrt{2}\mapsto
\hat{b}_+\;,
\end{equation}
so that a linearly polarized single-photon state is mapped onto
the state $|1\rangle_b$. This shows that a rotating half-wave
plate with linear polarized photons as input generates $b_+$
photons as output.

We can also take as input a linear superposition of modes with
orbital mode indices $m$ and $-m$, onto a mode converter rotating
at an angular frequency $\Omega/2$. The corresponding classical
mapping is
\begin{equation}
\vec{E}_{\omega,m,s}\exp(im\Omega t/2)\mapsto\vec{E}_{\omega -
m\Omega,-m,s}\exp(-im\Omega t/2)\;.
\end{equation}
There is, however, an ambiguity here: a mode converter will have
an input plane $z=z_i$ and a different output plane $z=z_o$. Since
between those planes modes with different frequencies will
diffract differently, a rotating mode converter works properly
only when $z_i$ and $z_o$ are sufficiently close for those
diffraction effects to be negligible. Assuming this is the case,
the quantum equivalent of the mapping by a rotating mode converter
is
\begin{equation}
\hat{a}_{\omega,m,s} \mapsto \hat{a}_{\omega-m\Omega,-m,s}\;.
\end{equation}
The mode operator corresponding to the superposition of two modes
with opposite orbital angular momentum $\pm m$ is mapped as
\begin{equation}
(\hat{a}_{\omega,m,s} + \hat{a}_{\omega,-m,s})/\sqrt{2} \mapsto
\hat{c}_+\;.
\end{equation}
A single photon in this superposition mode is therefore converted
into the single-photon state $|1\rangle_c$, with the single-photon
wave function $\vec{E}^c_+(\vec{r},t)$. Again, since this photon
is in an equal superposition of two states with orbital angular
momentum $\pm \hbar m$, its average orbital angular momentum is
zero, even though the mode pattern is rotating. This agrees with
the conclusions for a classical rotating field created by a
rotating mode inverter \cite{nien}.
\subsection{Measurements on single rotating photons}
Polarization measurements on single photons are certainly possible. If one wishes to distinguish, say, $x$ and $y$ polarized photons, all one needs is a polarizing beam splitter oriented such that $x$ and $y$ polarized photons exit at different output ports, and a subsequent photodetection event performs a (destructive) projective measurement. Similar projective measurements distinguishing different orbital angular momentum eigenstates of single photons are possible as well \cite{sortl}. In that case, too, one can construct a sorting device that splits an incoming stream of photons into different output channels with different (orthogonal) angular momentum states. A subsequent photodetection finalizes the projective measurement.

Can one do the same for rotating modes? That is, can one construct a similar sorting device for distinguishing, e.g., the modes $b_+$ and $b_-$ even for single photons? Using the preceding subsection it is easy to see one certainly can. Namely, one simply sends the input photon through a rotating half wave plate, after which a polarization measurement achieves the desired projective measurement. In order to distinguish modes $c_+$ and $c_-$ at the single-photon level one 
first sends the photon through a rotating mode converter, and subsequently one needs a sorting device that distinguishes between different Hermite-Gaussian modes. Fortunately, such devices, working at the single-photon level, exist as well \cite{HG}. 
\section{Rotation with angular momentum}\label{rot}
There is an alternative way of defining mode transformations. If
we consider the expression for the rotating polarization of the
mode of Eq.~(\ref{Eb}), then we see extra prefactors $\sin\theta$
and $\cos\theta$ appearing because of the (quantum) normalization
factor proportional to $\sqrt{\hbar\omega}$. Similar factors
appear in Eq.~(\ref{azi}) for the rotating transverse mode
pattern, for the same reason. In order to compensate for those
prefactors we replace the mode operators (\ref{rot1a}) by the
definition
\begin{eqnarray}\label{arot2}
\hat{g}_+&=& \sin\theta \hat{a}_{\omega+\Omega s,m,s}
+\cos\theta \hat{a}_{\omega-\Omega s,m,-s}, \nonumber\\
\hat{g}_-&=& \cos\theta \hat{a}_{\omega+\Omega s,m,s} -\sin\theta
\hat{a}_{\omega-\Omega s,m,-s}\;.
\end{eqnarray}
The compensation of the factor $\sqrt{\omega}$ works only for the
$`+'$ mode of the pair of modes (\ref{arot2}), but the companion
$'-'$ modes are necessary to make the redefinition unitary.

For the $'+'$ modes we get instead of (\ref{Eb}) the expression
for the electric-field operator
\begin{eqnarray}\label{Ef}
\vec{E}^g_{+}(t)\hat{g}_{+} &=&
\sqrt{\frac{\hbar\omega}{\epsilon_0 V}}\exp(-i\omega t)
\sin\theta\cos\theta
\exp(im\phi)F(\rho,z_0)\nonumber\\
&\times&\hat{g}_{+}\left[\vec{e}_+\exp(-i\Omega t) +
\vec{e}_-\exp(i\Omega t)
\right]\;.\nonumber\\
\end{eqnarray}
The last line now describes a rotating {\em linear} polarization
\begin{equation}\label{polf}
{\vec e}(t)={\vec e}_x \cos\Omega t + {\vec e}_y \sin\Omega t\;.
\end{equation}
A single photon from this mode now does possess a nonzero average
spin angular momentum, equal to
\begin{equation}
\langle \hat{S}_z\rangle=(\sin^2\theta-\cos^2\theta)\hbar=
-\hbar\Omega/\omega\;.
\end{equation}
This result is exact, not perturbative, even though typically we
do have $\Omega/\omega\ll 1$. The photon is in an unbalanced
superposition of states with angular momentum $+\hbar$ with
relative weight $\sin^2\theta$ and with angular momentum $-\hbar$
with weight $\cos^2\theta$. Similarly, the energy of a $g_+$
photon is
\begin{equation}
\langle E\rangle= \hbar\omega-\hbar\Omega^2/\omega\;.
\end{equation}

Of course we can define analogous modes with rotating transverse
mode patterns, while also compensating for the extra factors
$\sin\theta$ and $\cos\theta$ in Eq.~(\ref{azi}), by defining
\begin{eqnarray}\label{arot3}
\hat{h}_+&=& \sin\theta \hat{a}_{\omega+\Omega m,m,s}
+\cos\theta \hat{a}_{\omega-\Omega m,-m,s}\;, \nonumber\\
\hat{h}_-&=& \cos\theta \hat{a}_{\omega+\Omega m,m,s} -\sin\theta
\hat{a}_{\omega-\Omega m,-m,s}\;.
\end{eqnarray}
It is again only the $h_+$ modes for which the $\theta$-dependent
prefactors in the electric-field amplitude cancel. That is, the
electric field of such a mode rotates around the $z$ axis with the
same shape as in the non-rotating case $\Omega = 0$. A single
photon in the $h_+$ mode has an orbital angular momentum equal to
\begin{equation}
\langle \hat{L}_z\rangle= -\hbar m^2\Omega/\omega\;,
\end{equation}
and the energy is $\langle E\rangle= \hbar\omega-\hbar
m^2\Omega^2/\omega$. Thus for both modes $g_+$ and $h_+$the angular momentum is, perhaps counter-intuitively,
negative for a mode rotating in the positive direction around the
$z$ axis.

Interestingly, the $'-'$ modes that we were forced to define by
requiring unitarity also have a nice property: for these modes it
is the $\omega$-dependent prefactors in the expression for the
vector potential that cancel, rather than in the expression for
the electric field. Thus the $g_-$ mode describes a vector
potential whose direction rotates uniformly and without changing
length around the $z$ axis. Similarly, for the $h_-$ modes the
transverse mode pattern of the vector potential rotates around the
$z$ axis without changing shape. For single photons in the $'-'$
modes we find now that the angular momentum has the opposite value
as for the $'+'$ modes. Thus we have
\begin{equation}
\langle \hat{S}_z\rangle= \hbar \Omega/\omega\;,
\end{equation}
for a $g_-$ photon and
\begin{equation}
\langle \hat{L}_z\rangle= \hbar m^2 \Omega/\omega,
\end{equation}
for a $h_-$ photon. The energy of the photons is $\langle
E\rangle= \hbar\omega+\hbar\Omega^2/\omega$ and $\langle E\rangle=
\hbar\omega+\hbar m^2\Omega^2/\omega$, respectively. So here the
energy per photon is higher than $\hbar\omega$, while for the
$'+'$ modes it was lower by the same amount. As far as the authors
are aware, the $'-'$ modes have not been discussed before.

Of course, these other types of rotating photons can be generated
with rotating mode inverters by taking different superpositions as
input.

\section{Photons as seen from a rotating frame}\label{frame}
The modes we have constructed so far are ``rotating modes''. The
field modes satisfy the Maxwell equations or the paraxial
equations, and the mode operators satisfy the Heisenberg equations
of motion. It is useful to compare those modes to the modes we get
by applying a rotation operator of the form
\begin{equation}
\hat{R}(t)=\exp(i\hat{J}_z\Omega t)
\end{equation}
to non-rotating modes. The transformed mode operators
\begin{eqnarray}
\hat{a}'(t)=\hat{R}^\dagger \hat{a}_{\omega,m,s}(t)
\hat{R}=\exp(i\Omega(m+s) t )\hat{a}_{\omega,m,s}(t)
\end{eqnarray}
no longer satisfy the correct Heisenberg equations of motion for a
free field, because the unitary rotation operator depends on time.
Instead the mode operators and the corresponding field operators
describe modes as seen from a rotating frame, rotating at an
angular frequency $\Omega$ around the $z$ axis. One may easily
confuse operators like $\exp(i\Omega(m+s) t
)\hat{a}_{\omega,m,s}(t)$ with the similar operators
$\hat{a}_{\omega-\Omega(m+s),m,s}(t)$.  Their time dependence is
the same, but since the corresponding mode functions have
different frequencies, they satisfy different equations of motion,
and display different diffraction behavior.

On the other hand, the fact that a rotating beam of light can be
described by taking superpositions of modes with different values
of angular momentum and shifting the frequency in proportion to
the angular momentum can be explained by this very analogy. The
frequency shift can be seen as a rotational version of the Doppler
shift \cite{niend}.

Alternatively, the frequency shift proportional to angular momentum can be seen as a time-dependent
manifestation of a geometric phase \cite{kimble}, with the time
derivative of the phase equaling the frequency shift. For a
rotating polarization this shift arises from the Pancharatnam
phase \cite{pan}, for a rotating transverse mode pattern it arises
from the similar ``orbital'' geometric phase associated with mode
transformations \cite{enk}. The latter geometric phase was
measured recently in its time-independent form \cite{galvez} by
using the mode converter from \cite{beij}.

It may be that there is a deeper connection between angular
momentum of light and the various geometric phases of light:
according to Refs.~\cite{tiwari} the geometric phase arises only
when angular momentum is exchanged, and this was confirmed in
special cases in \cite{enk,banerjee}. A recent experiment
\cite{galvez2} indicates that this connection between angular
momentum exchange and the occurrence of a geometric phase may be
more general.

\section{Examples}\label{Quinfo}
We give here some examples of the use of rotating photons, mostly for quantum-information processing purposes. We do not claim rotating photons are superior than any other type of photons in this context, they just provide an interesting alternative.

We first discuss {\em wave-packets} of rotating photons.
\subsection{Wave-packets of rotating photons}
The modes defined so far are of infinite spatial and temporal
extent but are nevertheless fine for most theoretical purposes. In
practice a more useful definition of photons is in terms of
wavepackets that are of finite extent. They are easily defined in
the usual way: consider  a state of the form
    \begin{equation}
|1\rangle_F:=\int d\omega F(\omega) \hat{g}^\dagger_+(\omega)|{\rm
vac}\rangle,
    \end{equation}
where $F(\omega)$ is a normalized function
    \begin{equation}
\int d\omega |F(\omega)|^2 = \int dt |\tilde{F}(t)|^2=1,
    \end{equation}
with $\tilde{F}(t)$ the Fourier transform
    \begin{equation}
\tilde{F}(t)=\frac{1}{\sqrt{2 \pi}} \int d\omega
F(\omega)\exp(-i\omega t).
    \end{equation}
The corresponding electric field is then determined by
    \begin{equation}
\langle {\rm vac}|\hat{\vec{E}}(\vec{r},t)|1\rangle_F = \int
d\omega F(\omega) \vec{E}^g_{+,\omega}(t).
    \end{equation}
This electric field is proportional to
    \begin{equation}
\tilde{F}(t)[\vec{e}_x \cos\Omega t + \vec{e}_y \sin\Omega t].
    \end{equation}
If we assume that $F(\omega)$ has a finite width in frequency
space, then the probability of detecting such a rotating photon is
nonzero only in a finite time interval at each given position, and
nonzero in a finite spatial interval at any given time. The
rotating character of a wave-packet will be visible only if its
time duration is sufficiently long. Roughly speaking, if the
duration $\Delta t>2\pi/\Omega$, then a full rotation of the
polarization vector over $2\pi$ is included in the wave-packet. In
frequency space, this means the frequency spread should be at most
of order $\Omega$. We will investigate how such a wave-packet
interacts with a single atom in later in this Section.

\subsection{Entanglement between rotating modes}
A {\em rotating} equivalent of the well-known singlet state may be
defined as
\begin{equation}\label{RSf}
|RS\rangle:=\frac{\hat{g}^\dagger_+\otimes \hat{g}^\dagger_--
\hat{g}^\dagger_-\otimes \hat{g}^\dagger_+}{\sqrt{2}}|{\rm
vac}\rangle.
\end{equation}
Here, in a tensor product the first operator is meant to refer to a mode
located in $A$, the second to a mode located in a different
location $B$. This state $|RS\rangle$ then is entangled between
modes (or locations) $A$ and $B$. For example, this two-photon
state is anticorrelated with respect to (average) spin angular
momentum: if one detects one photon in the $g_+$ state, then it
has spin angular momentum equal to $-\hbar\Omega/\omega$. The other
photon in the other location is then necessarily in the $g_-$
state with the opposite average spin angular momentum.

But we can rewrite this same state in various different forms. For
example, we also have
\begin{equation}\label{RSb}
|RS\rangle=\frac{\hat{b}^\dagger_+\otimes \hat{b}^\dagger_--
\hat{b}^\dagger_-\otimes \hat{b}^\dagger_+}{\sqrt{2}}|{\rm
vac}\rangle.
\end{equation}
So if we measure whether one photon is in the $b_+$ or $b_-$ mode,
then the other photon will be found in the other mode. Here both
photons have zero spin angular momentum on average. 

Finally, we can
also write
\begin{equation}\label{RSa}
|RS\rangle=\frac{\hat{a}^\dagger_{\omega+\Omega,+}\otimes
\hat{a}^\dagger_{\omega-\Omega,-} -
\hat{a}^\dagger_{\omega-\Omega,-}\otimes
\hat{a}^\dagger_{\omega+\Omega,+}}{\sqrt{2}}|{\rm vac}\rangle
\end{equation}
in terms of our original mode operators. So if one photon is found
to have an energy $\hbar(\omega+\Omega)$ then the other one must
have an energy $\hbar(\omega-\Omega)$. And if one photon has
spin angular momentum $+\hbar$ then the other has $-\hbar$. 
The rotating singlet state $|RS\rangle$ thus has in common with the
standard nonrotating singlet state (at a single frequency
$\omega$) that the total angular momentum is zero, and that the
total energy is $2\hbar\omega$.

On the
other hand, if in the state $|RS\rangle$ one photon is found
rotating, then the other is rotating in the {\em same} direction,
even when the angular momenta are in fact opposite.
Moreover, when trying to violate Bell inequalities with the state $|RS\rangle$ the standard polarization measurements in fixed bases will not do. Instead one needs polarization measurements in co-rotating bases.

Similar conclusions hold for the orbital equivalent of $|RS\rangle$, defined by
\begin{eqnarray}\label{RSch}
|RSO\rangle&:=&\frac{\hat{c}^\dagger_+\otimes \hat{c}^\dagger_--
\hat{c}^\dagger_-\otimes \hat{c}^\dagger_+}{\sqrt{2}}|{\rm
vac}\rangle\nonumber\\
&=&
\frac{\hat{h}^\dagger_+\otimes \hat{h}^\dagger_--
\hat{h}^\dagger_-\otimes \hat{h}^\dagger_+}{\sqrt{2}}|{\rm
vac}\rangle.
\end{eqnarray}
\subsection{Quantum cryptography with rotating photons}
For quantum key distribution, in particular for the BB84 protocol
\cite{bb84,qkd}, one needs two mutually unbiased bases. One could
(in theory rather than in practice) use single photons and encode
information in polarization. One orthodox choice could be to use
single photons in the modes $a_{\omega,+}$ and $a_{\omega,-}$ as
one pair of orthogonal states (basis), corresponding to circular
polarization, and another pair of modes corresponding to linear
polarization, e.g. $(a_{\omega,+}\pm a_{\omega,-})/\sqrt{2}$.

In the present context, we could of course contemplate using one
pair of orthogonal {\em rotating} modes $b_{\pm}$ [with zero average spin angular momentum] and another
pair of nonrotating modes $a_{\omega+\Omega,+}$ and
$a_{\omega-\Omega,-}$, such that the overlaps between states from
different bases are equal to 50\%. In the latter set the two basis
states can be distinguished by a frequency measurement but
also by a polarization measurement. On the other hand, the former
two basis states can be distinguished by neither of those
measurements, and instead one could use a polarization measurement
in combination with a precise timing measurement ($b_+$ is a
time-shifted version of $b_-$). More precisely, one basis requires
a frequency measurement with an accuracy better than $\Omega$, the
other basis requires a timing measurement better than
$\pi/(4\Omega)$, thus revealing the complementarity between the
two bases. We could say that with the help of the (rotating)
polarization degree of freedom, the conjugate variables used for
this implementation of BB84 are time and frequency, rather than
noncommuting spin measurements $\sigma_z$ and $\sigma_x$ in the
orthodox implementation.

This implementation is related to, but different from, 
time-bin entangled photons for quantum key distribution
\cite{timebin}. In fact, the rotating version of the BB84 protocol
is equivalent to an entanglement-based protocol that makes use of
the rotating singlet state $|RS\rangle$ from the preceding
subsection. The fact we can write the same state in different
forms, namely (\ref{RSb}) and (\ref{RSa}), demonstrates this
explicitly.

The generalization to the orbital equivalent is obvious.
\subsection{Interference between rotating photons}
A typical quantum effect is the appearance of a dip at zero delay
in the number of coincidences of detector clicks behind a beam
splitter as a function of delay between two input photons that
enter the two input ports: the Hong-Ou-Mandel (HOM) dip
\cite{hom}. If we consider the interference of two, say, $b_+$
photons as a function of a time delay between them, then we find
the standard HOM curve is modulated by an extra time-dependent factor. Namely, we have
\begin{equation}
b_+(t+\tau)=\cos\Omega \tau b_+(t)-\sin\Omega\tau b_-(t),
\end{equation}
so that the modulation factor is simply
\begin{equation}
\cos^2\Omega \tau=\frac{1}{2}+\frac{1}{2}\cos2\Omega \tau.
\end{equation}
This implies the HOM curve will display an oscillation at a
frequency $2\Omega$ as a result of the rotating character of the
polarization. But,  of course, it can also be viewed as a beat
frequency between the $\omega\pm\Omega$ components of the modes.
Finally, we note the modulation factor is slightly different from
the overlap of the two
time-dependent polarizations,
\begin{equation}
|\vec{e}(t)\cdot \vec{e}^*(t+\tau)|^2=
\sin^4\theta+\cos^4\theta+2\sin^2\theta\cos^2\theta\cos(2\Omega \tau).
\end{equation}
\subsection{Single atoms and single rotating photons}
In principle {\em any} degree of freedom of a photon can be used
to encode information. One question that may arise in the present
context is whether the information about the rotating polarization
or rotating transverse mode pattern of a single photon can be
stored or processed or transferred to a different information
carrier. For that purpose let us consider here the interaction
between a single photon and a single atom.

One can certainly store the {\em fixed} (non-rotating)
polarization state of a photon in an atom by making use of the
selection rules (or, equivalently, angular momentum conservation).
Namely, starting off the atom in a particular magnetic sub level
one can make the atom (in principle at least) absorb one photon by
using stimulated Raman scattering by applying a, say, $\hat{z}$
polarized laser beam. The atom will end up in a particular ground
state picked out by the selection rules. For instance, if the
initial atomic state is chosen to be $|m_A=0\rangle_A$ (we use a subscript $A$ to indicate atomic degrees of freedom) then by
absorbing a $\sigma^{\pm}$ photon the atom ends up in the state
$|m_A=\pm1\rangle_A$ (assuming, of course, those levels exist in the
atom). A photon with a general polarization
$\alpha\sigma^++\beta\sigma^-$ will put the atom in the
corresponding coherent superposition $\alpha|_A+1\rangle+\beta
|-1\rangle_A$ in that case (where we use the symmetry of the
Clebsch-Gordan coefficients for the transitions used).

But how can one store a {\em rotating} polarization? One could use
one atom and create a rotating superposition $\alpha\exp(-i\Omega
t')|+1\rangle_A+\beta\exp(i\Omega t') |-1\rangle_A$ by lifting the
energy degeneracy of the two $m$-levels by applying an external
magnetic field. Here $t'=t-t_0$ where $t_0$ is determined by the
time the magnetic field was switched on (and by how the magnetic
field was turned on). This indeed is the quantum state of an atom
whose magnetic moment is rotating in time. For this to work one
needs {\em a priori} knowledge of the rotation frequency $\Omega$.
This is, of course, consistent with the observation that $\Omega$
is not really a quantum number, but a classical parameter labeling
different complete sets of modes (note we need to know the value
of $\omega$ of the incoming photon as well, to match it with a
resonant transition in the atom). In this way we can transfer one
qubit from a single rotating photon to the rotating magnetic
moment of a single atom. All that is needed is information on the
precise timing (to within order $\Omega^{-1}$) of the switching on
of the magnetic field. Recall that the difference between a $b_+$
photon and a $b_-$ photon is just a time shift by
$\tau=\pi/(2\Omega)$.

(We note there is in fact an experiment succeeding in mapping
certain quantum states of light onto an atomic ensemble, using
rotating spin states of atoms \cite{polzik}. However, what is
stored is a very different type of information, namely the
amplitude of a coherent state. The input state of the light field
is not rotating in the experiment of Ref.~\cite{polzik}.)

Let us consider the process of an atom absorbing a photon with
rotating polarization in more detail. We consider the case where
the external degrees of freedom of the atom can be treated
classically. An atom located at position $z=z_A$ not too far from
the focal plane $z=z_0$ starting in the state $|m_A=0\rangle_A$ will
turn into (using a subscript $F$ to indicate the field degrees of freedom)
\begin{eqnarray}
&&|1\rangle_F\otimes |0\rangle_A
\mapsto\nonumber\\
&&|0\rangle_F\otimes
\int d\tau \tilde{F}(\tau-z'/c)\exp(-i\omega(\tau-z'/c))\nonumber\\ &&\left[
P(\omega+\Omega)\exp(-i\Omega(\tau-z'/c))|+1\rangle_A
+\right.\nonumber\\
&&\left.P(\omega-\Omega)\exp(i\Omega(\tau-z'/c))|-1\rangle_A
\right]\nonumber\\
&&+\sqrt{P_{0}}|1\rangle_F\otimes|0\rangle_A,
\end{eqnarray}
where $z'=z_A-z_0$ is the position of the atom relative to the
focal plane $z=z_0$. We included here a probability amplitude for
the atom to absorb a photon at frequency $\omega$ of the form
$P(\omega)\propto (\omega-\omega_A-i\Gamma)^{-1}$, with $\omega_A$
the appropriate 2-photon Raman resonance frequency and $\Gamma$ an
(effective Raman) decay width. The last line gives the term
describing the case that the atom does not absorb the photon,
with probability $P_0$. Assuming $P_0\ll 1$ we see the atom will
end up in a roughly equal superposition of $|+1\rangle_A$ and
$|-1\rangle_A$, as if excited by a photon with a linear polarization
that is a weighted time average of the rotating polarization
passing by. Also note that the position of the atom is irrelevant:
the atom will see the same sequence of time-varying polarization
vectors pass by, no matter where it is (provided it is still near
the focal plane, so that diffraction effects can be neglected).

Now suppose the external degrees of freedom of the atom are
treated quantum mechanically, and the atom starts out in a pure
state of its center-of-mass motion. Because of energy and momentum
conservation the atom ends up in a state that displays correlation
(or even entanglement) between its external and internal degrees
of freedom. This way information about $\Omega$ can, in principle,
be stored. If we write the transformation the atom undergoes in
symbolic notation as
\begin{eqnarray}
&&|1\rangle_F\otimes |0\rangle_A\otimes |E\rangle_A
\mapsto\nonumber\\
&&|0\rangle_F\otimes
\int d\tau \tilde{F}(\tau-z'/c)\exp(-i\omega(\tau-z'/c))
\nonumber\\
&&\left[
P(\omega+\Omega)\exp(-i\Omega(\tau-z'/c))|+1\rangle_A\otimes |E+\Omega\rangle_A
+\right.\nonumber\\
&&\left.P(\omega-\Omega)\exp(i\Omega(\tau-z'/c))|-1\rangle_A\otimes |E-\Omega\rangle_A
\right]
\nonumber\\
&&+\sqrt{P_{0}}|1\rangle_F\otimes|0\rangle_A\otimes|E\rangle_A,
\end{eqnarray}
then the information about $\Omega$ is stored in the atom's external state provided, for instance,
 $\langle E-\Omega |E-\Omega'\rangle=0$ for $\Omega'\neq\Omega$.
After all, in that case a measurement on the external state of the
atom can reveal the value of $\Omega$. 

Under those same
circumstances the internal and external states of the atom are
entangled, since we also have  $\langle E-\Omega
|E+\Omega\rangle=0$, with maximum entanglement if
$|F(\omega+\Omega)|=|F(\omega-\Omega)|$. The more the two states
$|E-\Omega\rangle_A$ and $|E+\Omega\rangle_A$ overlap, the less
entangled internal and external states are, and the less
information is stored about the value of $\Omega$. In the extreme
case of perfect overlap, we are back to the classical case: for
example, the atom is in a quasi-classical coherent state of its
external motion, and the energy $\hbar\Omega$ is much smaller than
the average motional energy of the atom, so that the small shift
in energy is not detectable, and the external and internal degrees
of freedom remain uncorrelated.

For completeness, let us note that in order to transfer
information encoded in a rotating transverse mode profile, one
necessarily needs to use the external degrees of freedom. Indeed,
whereas spin angular momentum is coupled to the internal
(electronic) degrees of freedom of an atom, the orbital angular
momentum couples to the center-of-mass motion, as was shown
explicitly in Ref.~\cite{com}. Again, one needs {\em a priori}
knowledge of the value of $\Omega$, but then one can, in principle
at least if not in practice yet, transfer a bit of information
encoded in, say, the $c_{\pm}$ modes to an atom. This would,
however, be a much more involved experiment than related
experiments producing entanglement between modes of different
orbital angular momentum \cite{zeil,eliel}.

\section{Summary}\label{sum}
We developed a quantum theory of rotating photons for which either
the (linear or slightly elliptical)  polarization vector or the
transverse mode pattern rotates slowly around the propagation
($z$) direction. The rotational frequency $\Omega$ is independent
of the optical frequency $\omega$, and will typically be much
smaller than $\omega$. We found that there are, in each case,
{\em three} natural types of rotating photons: they can have spin angular
momentum $-\hbar \Omega/\omega$, $0$, or $+\hbar \Omega/\omega$,
if the photon has a rotating polarization, and an orbital angular momentum
$-m\hbar \Omega/\omega$, $0$, or $+m\hbar \Omega/\omega$ if the
photon has a rotating transverse mode pattern composed of modes
with orbital angular momenta $\pm m\hbar$. These three types of
rotating photons correspond to modes with a rotating unchanging
electric field vector, an {\em equal} superposition of opposite
angular momenta, and a rotating unchanging vector potential,
respectively. These photons should be distinguished  from nonrotating photons as
viewed from a frame rotating at $-\Omega$ around the $z$ axis. We
also defined propagating rotating wavepackets of finite duration,
giving a more realistic picture of what would be produced in an
experiment.

We then considered some examples of single-photon and two-photon
states illustrating properties of rotating photons. We defined a
rotating version of the standard singlet state, an entangled state
consisting of two photons with opposite angular momenta. A new
aspect of the rotating version of the state is that, if the
polarization of one photon is measured and found rotating in the
positive direction around the propagation axis, the other photon's
polarization necessarily rotates in the same direction. But if one
photon's polarization in that same state is found not to rotate,
then neither does the other. Thus whereas the angular momenta of
the two photons are anti-correlated, the sense of rotation is
correlated.

Rotating photons also allow one to use different conjugate
variables for, e.g., quantum key distribution. In particular,
instead of using nonorthogonal polarization states to encode
information,  time and frequency can be used as conjugate
variables, with the help of a rotating polarization. That is,
information can either be stored in the frequency of a single
nonrotating photon or in the timing of a rotating single photon,
but one cannot measure both properties at the same time.

We then verified whether information about the rotating character
of a single photon can be transferred to the state of a single
atom. We argued one can certainly create a rotating magnetic
moment matching the rotating polarization of an incoming photon in
a single atom, provided one has {\em a priori} knowledge of the
value of $\Omega$: one just uses a magnetic field to cause a
Zeeman shift equal to $\hbar\Omega$ and a superposition of
different Zeeman sublevels $|m=\pm 1\rangle$ then creates a
rotating magnetic moment, rotating at a frequency $\Omega$. The
fact that one needs classical information about the rotational
frequency is directly related to the fact that $\Omega$ is {\em
not} a quantum number, but rather a classical parameter that
labels different complete sets of (quantum) modes. With degenerate
magnetic sublevels the internal state of an atom stores merely a
time-average of the rotating polarization of an incoming photon.

We also considered  how the external motion of an atom can store
certain types of information encoded in a rotating photon. The
value of $\Omega$ can be stored, and, similarly, a transverse mode
pattern can be imprinted on the center-of-mass motion of a single
atom. Using the center-of-mass motion is perhaps not a very
practical idea, but we considered this case for the sake of
completeness.

Finally, it may be interesting to see if the concept of rotating
photons can prove useful in the field of time and frequency
standards. After all, the optical frequency $\omega$ of photons exploited
in such standards is complemented by a slowly rotating
polarization vector, which may be seen as an extra slow hand of a
fast clock. This does require that $\omega$ and the rotational frequency $\Omega$ be locked to each other by some process \cite{Jun}. It seems unlikely such a process exists in nature.

\end{document}